\documentclass[
    ,final            
  ]
  {aipproc}

\layoutstyle{6x9}

\begin{document}

\title{$\Lambda(1405)$ and kaonic few-body states in chiral dynamics}


\author{Daisuke Jido}{
  address={Yukawa Institute for Theoretical Physics, Kyoto University, Sakyo, Kyoto 606-8502, Japan}
}

%

\begin{abstract}
The present status of the $\Lambda(1405)$ structure study
in chiral dynamics is briefly reviewed. It turns out that the $\Lambda(1405)$ 
resonance can be described by hadronic dynamics. The idea of the hadronic
molecular state is extended to kaonic few-body states. It is concluded that,
due to the fact that $K$ and $N$ have similar coupling nature in $s$-wave 
$\bar K$ couplings, there are few-body quasibound states with kaons 
systematically just below the break-up thresholds, like $\bar KNN$, $\bar KKN$
and $\bar KKK$, as well as $\Lambda(1405)$ as a $\bar KN$ quasibound 
state and $f_{0}(980)$ and $a_{0}(980)$ as $\bar KK$. 
\end{abstract}

\maketitle


\section{Introduction}

Searching out effective constituents in the structure of hadrons is 
a clue to understand strongly interacting systems. Certainly, quarks and 
gluons are the fundamental constituents of hadrons, but the current quarks
appearing in the QCD Lagrangian cannot be effective degrees of freedom
to describe the hadron structure in simple and intuitive ways.  
Constituent quark models give a simple picture of the baryon structure,
reproducing the magnetic moments of the low-lying octet baryons by group
theoretical argument and also their masses by 
the Gell-Mann--Okubo mass formula, which is a consequence of the flavor 
SU(3) with its small breaking by the quark masses.
The success of the constituent quark models indicates that the symmetry of 
quark is realized in the baryon structure through the constituent quarks. 
In contrast, 
baryon resonances are decaying particles into mesons and a baryon by strong
interaction, and thus it is natural that baryon resonances may have large 
hadronic components apart from quark-originated components.
Therefore, for the investigation of the baryon resonance structure, the aspect 
of hadron dynamics is unavoidable to be considered. 
%
In this paper, 
we discuss hadronic resonance states in term of chiral dynamics. First,
we briefly review the present theoretical status of 
the $\Lambda(1405)$ resonance, which is now to be considered as one 
of the typical examples of meson-baryon quasibound states. 
Then we further develop the idea of the hadronic 
quasibound state into few-body hadronic systems with kaons.

\section{The structure of $\Lambda(1405)$}

The $\Lambda(1405)$ resonance is the lowest baryon with $J^{p}=(1/2)^{-}$.
It is not easy to explain the light $\Lambda(1405)$ mass 
and the $LS$ splitting against $\Lambda(1520)$ in simple quark models, once
one determines the model parameters in the nucleon sector. 
$\Lambda(1405)$ is located in a 100 MeV window 
between the $\pi\Sigma$ and $\bar KN$ thresholds and 
the decay width of $\Lambda(1405)$ is given by the open $\pi\Sigma$ channel.
Thus, for the study of the $\Lambda(1405)$ structure,
one needs dynamical description with coupled channels, at least, with the $\bar KN$
and $\pi \Sigma$ channels.

One of the powerful theoretical frameworks to describe hadronic resonances
in hadron dynamics is the coupled-channels approach based on chiral dynamics,
so-called chiral unitary model, in which the resonances are described in 
hadron-hadron scattering~\cite{Kaiser:1995eg,Oller:1997ti}. The elementary
interactions of the scattering system are introduced by the chiral effective 
Lagrangian, and scattering equation is solved in a way to keep the $s$-channel 
unitarity. For the $s$-wave meson-baryon scattering with $S=-1$ and $Q=0$, 
this approach successfully reproduces the total and differential cross sections of 
$K^{-}p$ scatterings, and the $\Lambda(1405)$ resonance is obtained as a 
dynamically generated object without introducing explicit pole 
terms~\cite{Kaiser:1995eg,ChUM,Oller:2000fj}.
$\Lambda(1405)$ was found also to be an almost purely 
hadronic object~\cite{Hyodo:2008xr}.
After reproducing the observed scattering quantities, the chiral unitary model can be
used to investigate the structure and properties of the dynamically generated
resonances. The mass and width is obtained as the pole position of the scattering 
amplitude and the coupling nature of the resonance is extracted from the residues
of the pole.

\begin{figure}[tb]
  \includegraphics[width=0.4\textwidth,bb=0 0 516 363]{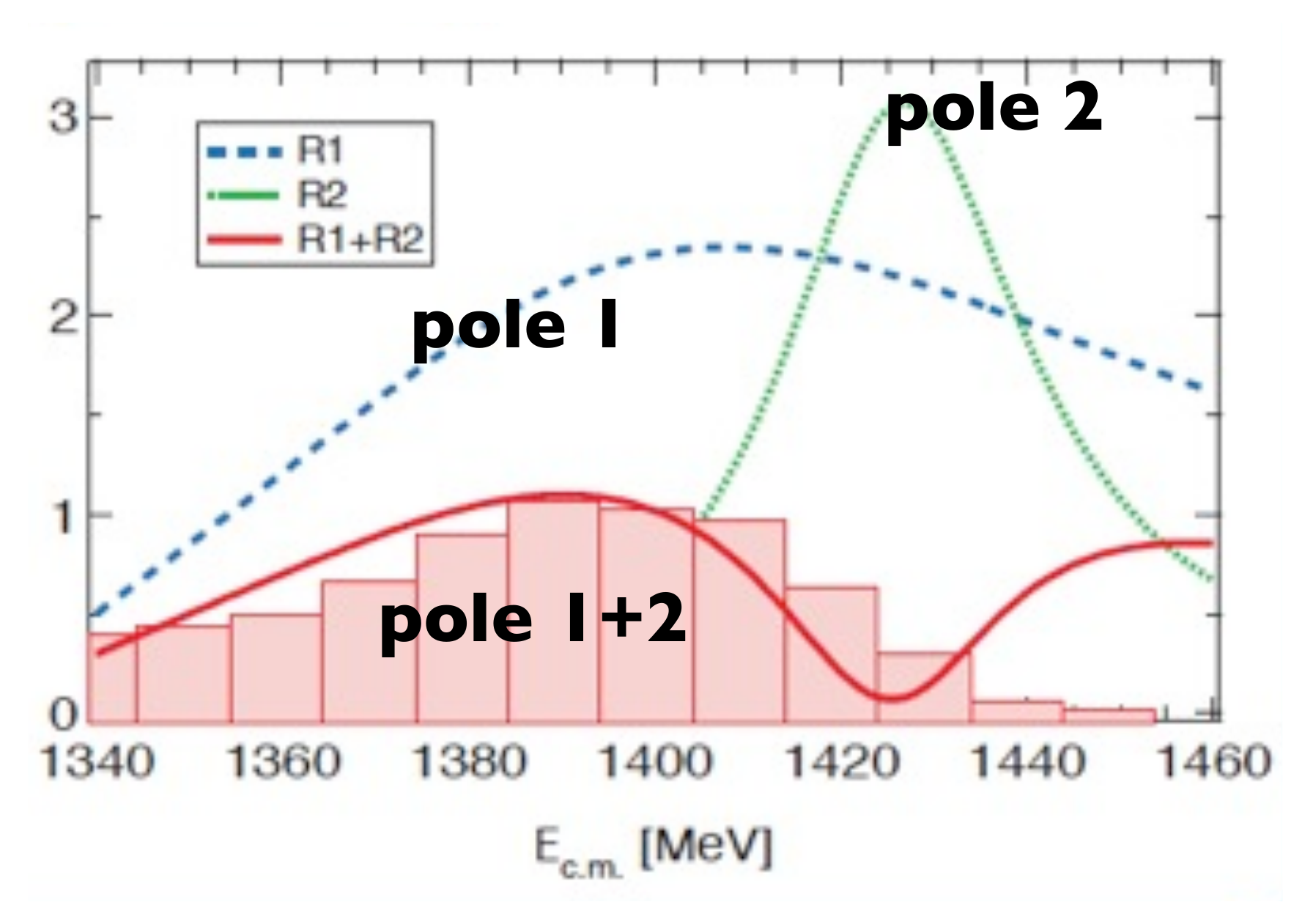}
  \includegraphics[width=0.4\textwidth,bb=0 0 516 363]{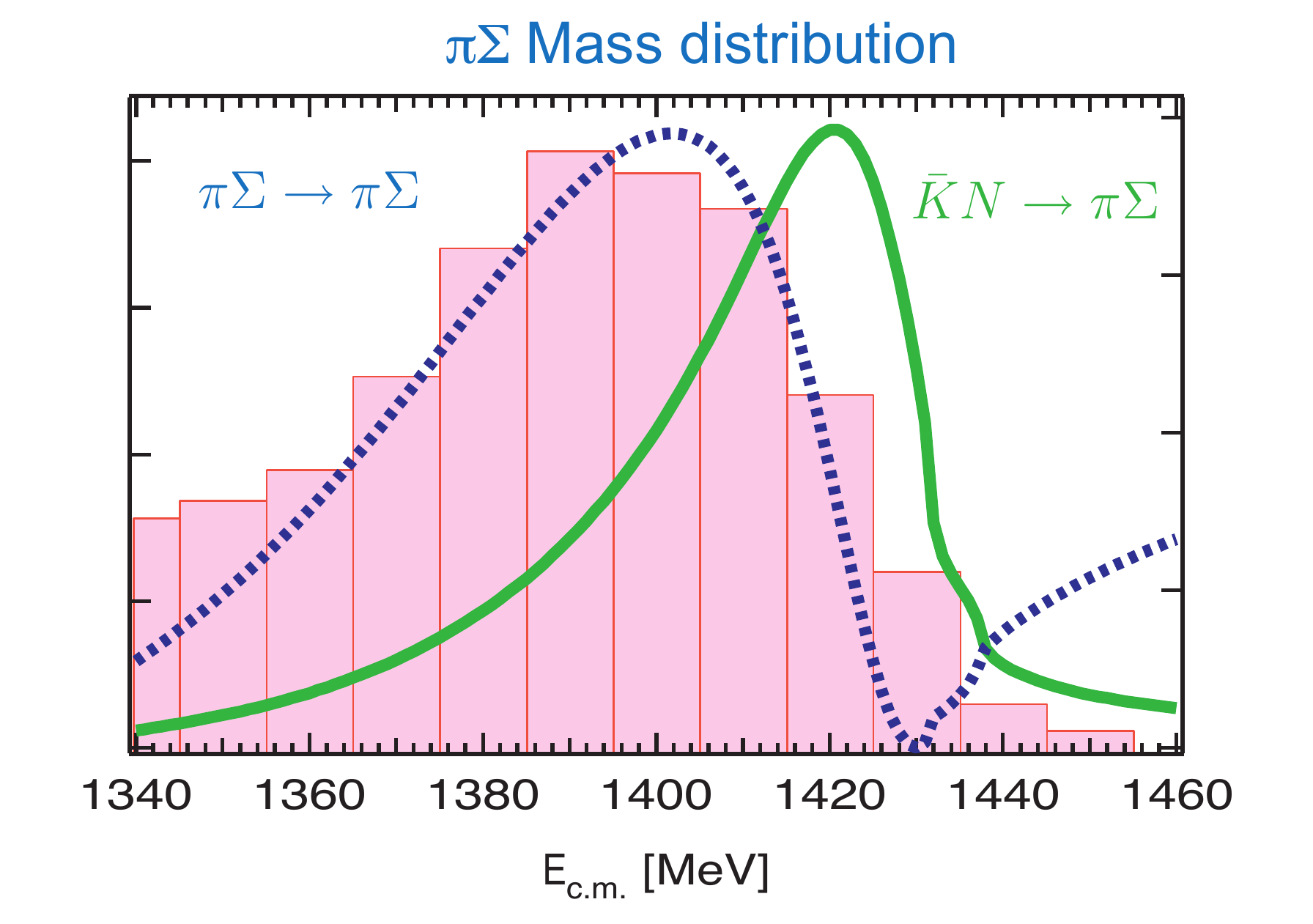}
  \caption{Mass spectra of $\Lambda(1405)$ in arbitrary units~\cite{Jido:2003cb}. 
  The left panel shows the $\pi\Sigma$ invariant mass spectrum (red line) of 
  $\pi\Sigma \to \pi\Sigma$ with $I=0$ calculated by the chiral unitary model 
  together with the separated contributions of the lower and higher poles of 
  $\Lambda(1405)$. The right panel shows a comparison of the $\pi\Sigma$ 
  spectra of $\pi\Sigma \to \pi\Sigma$ and $\bar KN \to \pi\Sigma$. (The heights are 
  adjusted.) }
  \label{fig:spec}
\end{figure}

One of the important consequence of the chiral unitary model for $\Lambda(1405)$ 
is that the observed spectrum of $\Lambda(1405)$ is composed by two resonance 
states having different coupling nature~\cite{Jido:2003cb}. 
The two poles for $\Lambda(1405)$ in chiral dynamics was found 
first in Ref.~\cite{Oller:2000fj} and later its significance was discussed 
in Ref.~\cite{Jido:2003cb}.
One of the pole is located at $1390 - 66 i$ MeV with a strong coupling to 
the $\pi\Sigma$ channel, while the other is at $1426 - 16i$ MeV with a dominant 
coupling to the $\bar KN$ channel~\cite{Jido:2003cb,Jido:2002yz}. 
Although these two states appear definitely different positions in the complex
plane, due to their closer masses and decay widths, their contributions to 
the $\Lambda(1405)$ spectrum seen on the real axis are entangled and these 
two states cannot be seen separately in the $\Lambda(1405)$ spectrum. 
What one can observe in experiments is interference of these two states (see the 
left panel of Fig.~\ref{fig:spec}). 
As a result of the presence of the two states with the different nature of 
the meson-baryon couplings,
the $\Lambda(1405)$ spectrum depends on the initial channel of 
the $\Lambda(1405)$ production~\cite{Jido:2003cb}. For example,
if the $\Lambda(1405)$ resonance is created by the $\bar KN$ channel, 
the higher pole gives more contribution to the spectrum than the lower pole. 
Consequently the resonance peak in the $\pi\Sigma$ spectrum appears 
around 1420 MeV, as shown in the right panel of Fig.~\ref{fig:spec}. 
This means that, since initial wight of the $\pi\Sigma$ and $\bar KN$
contributions to the $\Lambda(1405)$ production is different in each reaction 
mechanism, the peak position 
can be different in each reaction. Thus, when one investigates the 
$\Lambda(1405)$ properties from scattering experiments, it is extremely 
important to pin down the production mechanism of $\Lambda(1405)$
in each experiment.

The evidence of the double pole structure of $\Lambda(1405)$ is given 
by observing the peak position of the $\Lambda(1405)$ spectrum 
initiated by the $\bar KN$ channel, which will be at around 1420 MeV 
and be different from the resonance position observed in other experiments. 
Since $\Lambda(1405)$ is sitting below the $\bar KN$ threshold, one cannot 
produce $\Lambda(1405)$ directly by the $\bar KN$ channel. 
A recent work has shown that $K^{-}d \to \Lambda(1405) n$ is a good 
reaction~\cite{Jido:2009jf},
in which kaon brings strangeness  into the system and 
$\pi\Sigma$ cannot be the initial state of $\Lambda(1405)$ production. 
The theoretical calculation~\cite{Jido:2009jf} 
has also shown that the $\Lambda(1405)$ resonance is produced mostly 
in backward angles as a consequence of dominant contributions from the two-step
reaction,  and the $\Lambda(1405)$ spectrum peaking at 1420 MeV is 
consistent with an old bubble chamber experiment of 
$K^{-}d \to \pi^{+}\Sigma^{-}n$~\cite{Braun:1977wd}. Further detailed 
experiments will be performed at J-PARC~\cite{Noumi:J-PARC-E31} and 
DAFNE. It is also worth to mention that, in $K^{-}d \to Y\pi n$ reaction,
since the resonance position of the $\Lambda(1405)$ produced by $\bar KN$ 
will be at 1420 MeV with a narrower width, one could have a chance to 
observe $\Sigma(1385)$ and $\Lambda(1405)$ as separated peaks
in the missing mass spectrum of the emitted neutron~\cite{Yamagata:preparation}.

The reason that there are two poles around the $\Lambda(1405)$ energies is that
the elementary meson-baryon interaction has two attractive channels,
single and octet in the SU(3) flavor language or alternatively $\bar KN$ and 
$\pi\Sigma$ in the particle basis. It has turned out in Ref.~\cite{Hyodo:2007jq} 
that $\Lambda(1405)$ is essentially described by $\bar KN$ and 
$\pi\Sigma$, and the $\eta\Lambda$ and $K\Xi$ channels give 
minor contributions to the $\Lambda(1405)$ poles. In the single channel 
calculation there are a $\bar KN$ bound state and a $\pi\Sigma$ resonance 
in the $\bar KN$ and $\pi\Sigma$ channels, respectively, while 
with the $\pi\Sigma$-$\bar KN$ channel coupling, the $\bar KN$ 
bound state obtains its decay width to the $\pi\Sigma$ channel.
Therefore, we conclude that the $\Lambda(1405)$ resonance is described 
by the $\bar KN$ bound state and the $\pi\Sigma$ strong correlation. 
This is the meaning of the double pole structure of $\Lambda(1405)$. 
There is minor model dependence of the $\Lambda(1405)$ pole positions within the chiral 
unitary approach~\cite{Hyodo:2007jq}. The higher pole, which 
dominantly couples to $\bar KN$, has less model dependence in the pole position, 
located around 1420 MeV, because it is constrained by well-observed $\bar KN$ 
scattering data, while the lower pole strongly coupling to $\pi\Sigma$ has strong 
model-dependence due to lack of $\pi\Sigma$ scattering data. This implies 
that the $\bar KN$ scattering data alone cannot determine the full structure 
of $\Lambda(1405)$, and data for the $\pi\Sigma$ scattering, such as the 
scattering length and effective range, are necessary to investigate the structure of 
$\Lambda(1405)$ further~\cite{Ikeda:2011dx}.

\begin{figure}[tb]
  \includegraphics[width=0.70\textwidth,bb=0 0 552.79 170.08]{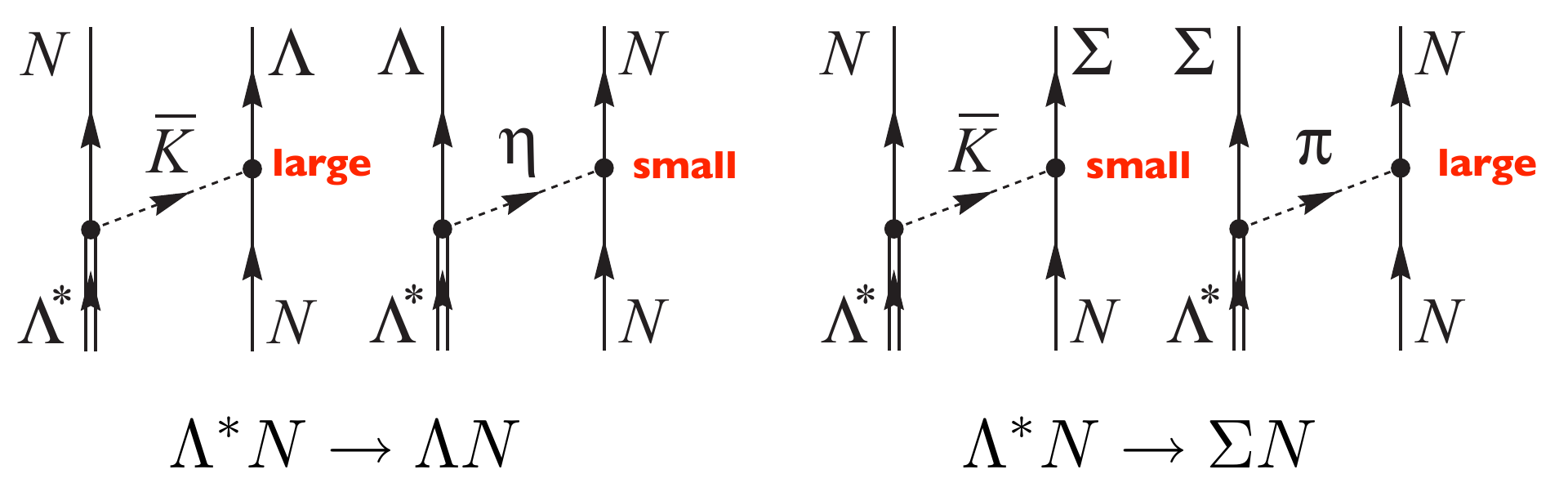}
  \caption{Diagrams for $\Lambda^{*}N \to YN$ transition in one meson-exchange 
  model~\cite{Sekihara:2009yk}. The left two diagrams is for 
  $\Lambda^{*}N \to \Lambda N$ and, the right two for $\Lambda^{*}N \to \Sigma N$.}
  \label{fig:dia}
\end{figure}

Dominance of $\bar KN$ in the $\Lambda(1405)$ resonance could be seen
in the nonmesonic decay of $\Lambda(1405)$ in nuclear matter~\cite{Sekihara:2009yk}. 
Figure~\ref{fig:dia} shows possible one-meson exchange diagrams for the 
$\Lambda^{*}N \to YN$ ($Y=\Lambda$ or $\Sigma$) transition. Since, according to 
the SU(3) flavor relation of the meson-baryon couplings, the $\eta NN$ and 
$\bar K N\Sigma$ couplings are small, the $\bar K$ ($\pi$) exchange is dominated  
in the $\Lambda^{*} N \to \Lambda N$ ($\Lambda^{*} N \to \Sigma N$) transition. 
Therefore, the ratio of the transition rates of $\Lambda^{*}N$ to $\Lambda N$ and
$\Sigma N$ is strongly sensitive to the $\Lambda^{*}$ coupling strengths
to $\bar KN$ and $\pi\Sigma$. 
The form factor of $\Lambda(1405)$ was also calculated within the chiral unitary approach 
using the chiral effective theory for the couplings of the external currents with 
the hadronic constituents~\cite{Jido:2002yz,Sekihara:2008qk}.
The density distributions and mean-squared radii were calculated 
from the form factor~\cite{Sekihara:2008qk}.
The electric radus of $\Lambda(1405)$ at the pole position is obtained as a complex
number $-0.157 + 0.238i$ fm$^{2}$, whose modules is much larger than that of 
neutron. This implies that kaon spreads spatially in wide range around nucleon.
It is also interesting seeing the $\Lambda(1405)$ resonance in the SU(3) 
limit~\cite{Jido:2003cb},
to see the role of the flavor symmetry in dynamically generated resonance. 
In a SU(3) flavor limit in which the octet mesons and baryons have averaged masses,
there are two bound states with 70 MeV and 5 MeV binding energies in 
the flavor single and octet channels, respectively. These binding energies 
are smaller than typical SU(3) breaking scale, $\sim 150$ MeV,
and therefore the SU(3) breaking effects should be larger than the low lying 
hadrons. This is a different aspect from constituent quark models.


\section{Kaonic few-body systems}

As we discussed above, it is most probable that $\Lambda(1405)$ is a 
dynamically generated resonance of a meson and a baryon, and especially one 
of the states can be a quasi-bound state of $\bar KN$. 
Such a quasibound state is called as hadronic molecular state. 
Hadronic molecular state is a (quasi) bound system composed
of hadronic constituents which keep their identity as they are in isolated systems,
appearing just below the threshold of break-up into the constituent hadrons. 
Driving force to make hadronic molecular states is hadronic interaction 
rather than inter-quark dynamics and confinement force. Thus, inter-hadron 
distances inside the hadronic molecular states are larger than the typical size 
of the low-lying hadrons which is characterized by the quark confinement range. 
Nucleus, which is a bound system of baryons, is also classified into this category.

Here let us consider hadronic molecular states with kaon and nucleon constituents.
For hadronic molecular states, pion has a too light mass to form bound states 
with other hadrons by hadronic interaction, since the pion kinetic energy in a confined 
system by hadronic interaction overcomes attractive potential energy. 
In contrast, kaon plays a unique role in hadronic molecular state due to 
Nambu-Goldstone boson nature and its heavier mass. Chiral effective theory suggests
$s$-wave attraction in the $\bar KN$ and $\bar KK$ channels, which is enough strong 
to form two-body quasibound states. The mass of kaon is so moderately heavy 
that kaon kinetic energy in hadronic bound systems can be smaller.
In kaonic few-body systems, hadronic molecular states are unavoidably resonances 
decaying into pionic channels.

\begin{figure}
  \includegraphics[width=0.65\textwidth,bb=0 0 729 514]{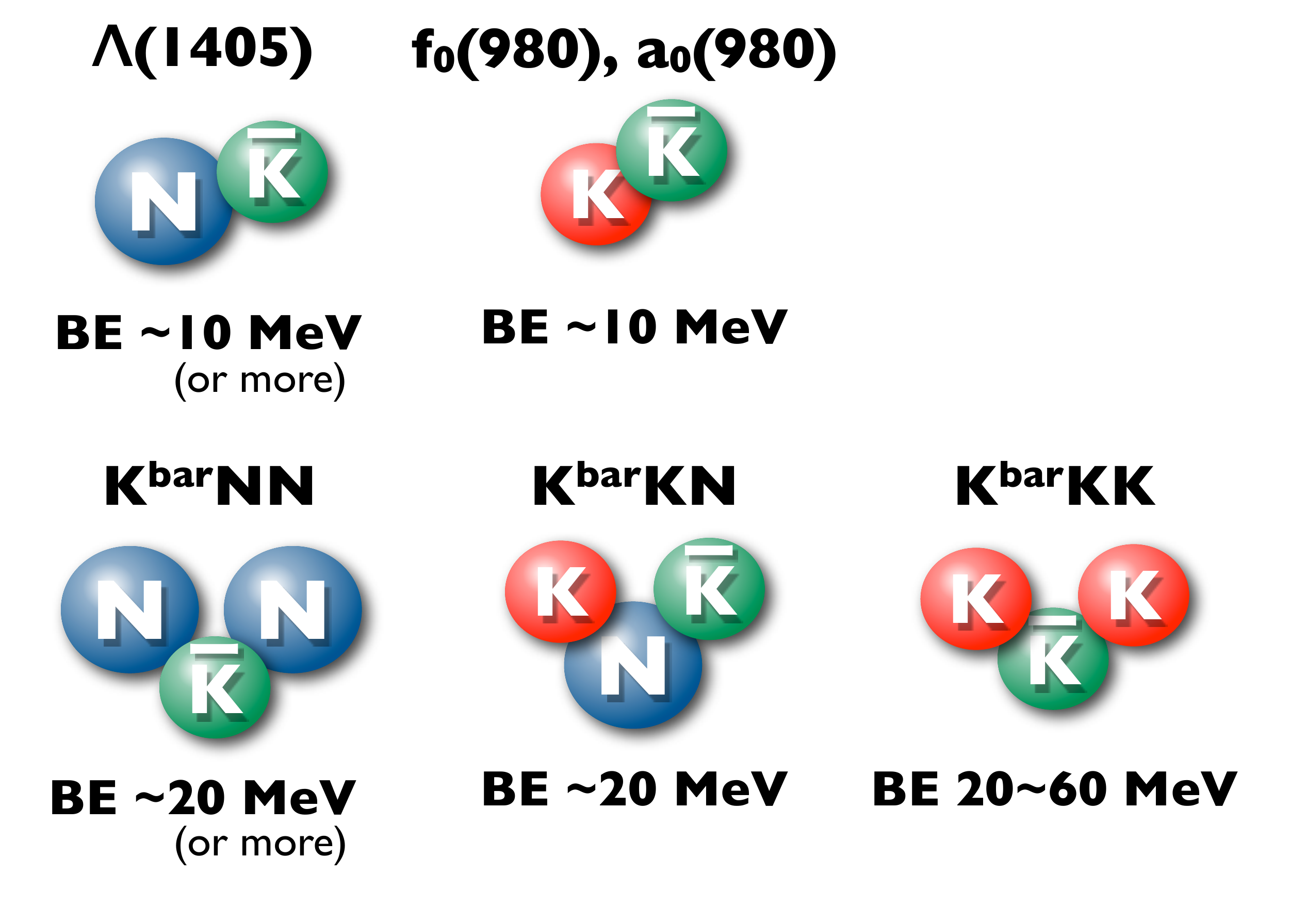}
  \caption{Family of kaonic few-body states. The binding energies are calculated in chiral dynamics.}
  \label{fig:family}
\end{figure}

In chiral dynamics the fundamental interaction is given by chiral effective theory,
and for $s$-wave interaction, the Tomozawa-Weinberg interaction is a driving force
of the hadronic molecular system. It is an interesting fact that the strength of the 
Tomozawa-Weinberg interaction is given by SU(3) flavor symmetry and 
$K$ and $N$ are classified into the same state vector in the octet representation. 
Therefore, 
the fundamental
interactions in $s$-wave are very similar in the $\bar KK$ and $\bar KN$ channel.
Consequently in these channels with $I=0$, there are quasibound states of 
$\bar KK$ and $\bar KN$ with a dozen MeV binding energy. 
This similarity between $K$ and $N$ is responsible for systematics of three-body
kaonic systems, $\bar KNN$, $\bar KKN$ and $\bar KKK$, as shown 
Fig.~\ref{fig:family}.

The celebrated $\bar KNN$ state with $I=1/2$ can be classified into the kaonic 
few-body family. This state was originally suggested 
by~\cite{Nogami:1963yf,Akaishi:2002bg}, 
and recent theoretical investigations 
shows that the $\bar KNN$ system is bound with a large width in spite of 
discrepancy between theoretical predictions 
of the mass and width. A study based on chiral dynamics predicts smaller binding
energy around 20 MeV~\cite{Dote:2008hw}. For further investigation 
it is extremely important to understand coupled-channel effects of 
$\pi\Sigma$ to the $\bar KNN$ system. If it gives significant contribution, 
especially it is probable for a larger binding case, one should treat $\pi\Sigma$ 
as a dynamically active channel. For this purpose, one needs
more precise data on two-body $\pi\Sigma$ scatterings~\cite{Ikeda:2011dx}.

The $\bar KKN$ and $\bar K \bar KN$ states with $I=1/2$ and $J^{p}=1/2^{+}$, 
which are $N^{*}$ and $\Xi^{*}$ resonances, respectively, were studied 
first in Refs.~\cite{Jido:2008kp,KanadaEnyo:2008wm} 
with a single-channel non-relativistic potential model. The $\bar KKN$
system was found to be bound with 20 MeV binding 
energy~\cite{Jido:2008kp}, and later  was investigated in a more 
sophisticate calculation~\cite{MartinezTorres:2008kh,MartinezTorres:2010zv} 
based on a coupled-channels Faddeev method developed in 
Refs.~\cite{MartinezTorres:2007sr},
in which a very similar state to one obtained in the potential model was found. 
Recently the $\bar KKN$ state was found also in a fixed center approximation
of three-body Faddeev calculation~\cite{Xie:2010ig}. The $\bar KKN$ state is
essentially described by a coexistence of $K \Lambda(1405)$ and 
$f_{0}(980)N$~\cite{Jido:2008kp}. An experimental search for  $\bar KKN$ 
was discussed in Ref.~\cite{MartinezTorres:2009cw}.

The $\bar KKK$ state with $I=1/2$ and $J^{p}= 0^{-}$, being an excited state of kaon,
was studied in a two-body $f_{0}K$ and $a_{0}K$ dynamics~\cite{Albaladejo:2010tj},
in the three-body Faddeev calculation~\cite{Torres:2011jt} and in the non-relativistic 
potential model~\cite{Torres:2011jt}. The three-body Faddeev calculation was done 
in coupled-channels of $K \bar KK$, $K \pi\pi$ and $K\pi\eta$ and a resonance 
state was found at 1420 MeV, while the potential model suggested a quasi bound
state with a binding energy 20 MeV. This state is essentially described by 
the $\bar KKK$ single channel and its configuration is found to be mostly $f_{0}K$. 
Experimentally, Particle Data Group tells that there is a excited kaon around 1460 
MeV observed in $K\pi\pi$ partial-wave analysis, although it is omitted 
from the summary table. 

In the potential model calculations of the $\bar KKN$ and $\bar KKK$ states,
it was found that the root mean-squared radii of these systems are as
large as 1.7 fm, which are similar with the radius of $^{4}$He.
The inter-hadron distances are comparable with an average 
nucleon-nucleon distance in nuclei. It was also found that the two-body
subsystems inside the three-body bound state keep their properties in 
isolated two-body systems. These features are caused by weakly binding of the 
three hadrons, for which the $KN$ and $KK$ repulsive interaction plays 
an important role to form hadronic molecular states.

\section{Summary and outlook}

We have discussed the structure of the $\Lambda(1405)$ resonance
in chiral dynamics. We have found that the $\Lambda(1405)$ resonance
can be described by meson-baryon dynamics. Due to the presence
of two attractive channels in fundamental meson-baryon interactions
the observed $\Lambda(1405)$ resonance is composed by two pole states.
One of the states is a quasi-bound state of $\bar KN$ located at around
1420 MeV and dominantly couples to the $\bar KN$ channel. Thus, this
is the relevant resonance for the kaon-nucleus interaction.
This finding is completely a consequence of chiral dynamics with constraint 
from the $K^{-}p$ scattering data, not a theoretical ansatz nor one putting 
by hand, unlike Ref.~\cite{Akaishi:2002bg} in which the $\Lambda(1405)$
position, 1405 MeV, is an input.
The double pole structure of $\Lambda(1405)$ can be confirmed experimentally 
by observing $K^{-}d \to \Lambda(1405) n$, in which 
$\Lambda(1405)$ is produced by the $\bar KN$ channel
and the  peak position appears around 1420~MeV.

The idea that $\Lambda(1405)$ is a quasibound state of $\bar KN$
can be extended to further few-body states with kaons like $\bar KNN$,
$\bar KKN$ and $\bar KKK$ having dozens MeV binding energy. 
In these states, a unique role of kaon is responsible for the systematics 
of the few-body kaonic states. Kaon has a half mass of nucleon and 
a very similar coupling nature to nucleon in the $s$-wave chiral interaction. 
This leads to weakly bound systems within the hadronic interaction range.
The hadronic molecular state is a concept of weakly binding systems 
of hadron constituents. If a resonance state has a large binding energy
measured from the break-up threshold, coupled-channel effects, like
$\pi\Sigma$ against $\bar KN$, and/or shorter range quark dynamics
may be important for the resonance state. In such a case the hadronic
molecular picture is broken down, and one should take into account 
the coupled channels contributions and quark dynamics for the study 
of the structure.

For some specific hadronic resonance, hadrons themselves can be 
effective constituents. 
The hadronic molecular configuration is a complemental picture 
of hadron structure to constituent quarks, which successfully 
describe the structure of the low-lying baryons in a simple way. 
Strong diquark configurations inside hadrons can
be effective constituents~\cite{Kim:2011ut}, 
and mixture of hadronic molecular and quark originated states
is also probable in some hadronic resonances~\cite{Nagahiro:2011jn}.
The hadronic molecular state has a larger
spacial size than the typical low-lying hadrons. 
In heavy ion collision, coalescence of hadrons to produce 
loosely bound hadronic molecular systems is more probable 
than quark coalescence for compact multi-quark systems~\cite{Cho:2010db}. 
Thus, one could extract the structure of hadrons by observing  
the production rate in heavy ion collisions.


\begin{theacknowledgments}
The auther would like to thank his collaborators for their collaborations on 
the works presented here.   
This work was supported by the Grant-in-Aid for Scientific Research from 
MEXT and JSPS (Nos.\ 22740161, 22105507).
This work was done in part under the Yukawa International Program for Quark-hadron Sciences (YIPQS).
\end{theacknowledgments}



%

\begin{thebibliography}{35}

\bibitem[Kaiser et~al.(1995)]{Kaiser:1995eg}
N.~Kaiser, P.~B. Siegel, and W.~Weise, \emph{Nucl. Phys.} \textbf{A594},
  325 (1995). 

\bibitem[Oller and Oset(1997)]{Oller:1997ti}
J.~A. Oller, and E.~Oset, \emph{Nucl. Phys.} \textbf{A620}, 438 (1997).

\bibitem[XXX(2000)]{ChUM}
E.~Oset, and A.~Ramos, \emph{Nucl. Phys.} \textbf{A635}, 99 (1998);
%
M.~F.~M. Lutz, and E.~E. Kolomeitsev, \emph{Nucl. Phys.} \textbf{A700}, 193 (2002); 
%
D.~Jido, E.~Oset, and A.~Ramos, \emph{Phys.Rev.} \textbf{C66}, 055203 (2002);
%
T.~Hyodo, S.I. Nam, D.~Jido, and A.~Hosaka, \emph{Phys. Rev.} \textbf{C68},
018201 (2003);
%
\emph{Prog. Theor. Phys.} \textbf{112}, 73 (2004).

\bibitem[Oller et~al.(2001)]{Oller:2000fj}
  J.~A.~Oller and U.~G.~Meissner,
  \emph{Phys. Lett.}  \textbf{B500}, 263 (2001).

\bibitem[Hyodo et~al.(2008)]{Hyodo:2008xr}
T.~Hyodo, D.~Jido, and A.~Hosaka, \emph{Phys. Rev.} \textbf{C78}, 025203 (2008);
See also T.~Hyodo {\it et al,}, in this proceedings.


\bibitem[Jido et~al.(2003)]{Jido:2003cb}
D.~Jido, J.~A. Oller, E.~Oset, A.~Ramos, and U.~G. Meissner, \emph{Nucl. Phys.}
  \textbf{A725}, 181 (2003). 

\bibitem[Jido et~al.(2002)]{Jido:2002yz}
D. Jido, A. Hosaka, J. C. Nacher, E. Oset and A. Ramos,
\emph{Phys. Rev.} \textbf{C66}, 025203 (2002).


\bibitem[Jido et~al.(2009)]{Jido:2009jf}
D.~Jido, E.~Oset, and T.~Sekihara, \emph{Eur. Phys. J.} \textbf{A42}, 257
  (2009);
%
 arXiv:1008.4423 [nucl-th].
 
\bibitem[Braun et~al.(1977)]{Braun:1977wd}
O.~Braun, {\it et~al.}, \emph{Nucl. Phys.} \textbf{B129}, 1 (1977).

\bibitem[Noumi et~al.(????)]{Noumi:J-PARC-E31}
M.~Noumi, et~al.,  J-PARC proposal E31 
  ``Spectroscopic study of hyperon resonances below $\bar KN$ threshold 
  via the $(K^{-},n)$ reaction on Deuteron'' (2009);
  S.~Enomoto, talk given in this conference. 

\bibitem[Yamagata-Sekihara et~al.(2011)]{Yamagata:preparation}
J.~Yamagata-Sekihara, T.~Sekihara, and D.~Jido, in preparation.

\bibitem[Hyodo and Weise(2008)]{Hyodo:2007jq}
T.~Hyodo, and W.~Weise, \emph{Phys. Rev.} \textbf{C77}, 035204 (2008).

\bibitem[Ikeda et~al.(2011)]{Ikeda:2011dx}
Y.~Ikeda, T.~Hyodo, D.~Jido, H.~Kamano, T.~Sato, and K.~Yazaki,
arXiv:1101.5190 [nucl-th].

\bibitem[Sekihara et~al.(2009)]{Sekihara:2009yk}
T.~Sekihara, D.~Jido, and Y.~Kanada-En'yo, \emph{Phys. Rev.} \textbf{C79},
  062201 (2009).
 
\bibitem[Sekihara et~al.(2008)]{Sekihara:2008qk}
T.~Sekihara, T.~Hyodo, and D.~Jido, \emph{Phys.Lett.} \textbf{B669}, 133
  (2008);
 %
arXiv:1012.3232 [nucl-th];
T.~Seki\-hara {\it et al.}, in this proceedins.
 
\bibitem[Nogami(1963)]{Nogami:1963yf}
Y.~Nogami, \emph{Phys. Lett.} \textbf{7}, 288 (1963);
%
See also for the present status, A.~Dot\'e, talk in this conference. 

\bibitem[Akaishi and Yamazaki(2002)]{Akaishi:2002bg}
Y.~Akaishi, and T.~Yamazaki, \emph{Phys. Rev.} \textbf{C65}, 044005 (2002);

\bibitem[Dote et~al.(2009)]{Dote:2008hw}
A.~Dote, T.~Hyodo, and W.~Weise, \emph{Phys. Rev.} \textbf{C79}, 014003 (2009).

\bibitem[Jido and Kanada-En'yo(2008)]{Jido:2008kp}
D.~Jido, and Y.~Kanada-En'yo, \emph{Phys. Rev.} \textbf{C78}, 035203 (2008).

\bibitem[Kanada-En'yo and Jido(2008)]{KanadaEnyo:2008wm}
Y.~Kanada-En'yo, and D.~Jido, \emph{Phys. Rev. C} \textbf{78}, 025212 (2008).

\bibitem[Martinez~Torres et~al.(2009{\natexlab{a}})]{MartinezTorres:2008kh}
A.~Martinez~Torres, K.~P. Khemchandani, and E.~Oset, \emph{Phys. Rev.}
  \textbf{C79}, 065207 (2009{\natexlab{a}}). 

\bibitem[Martinez~Torres and Jido(2010)]{MartinezTorres:2010zv}
A.~Martinez~Torres, and D.~Jido, \emph{Phys. Rev.} \textbf{C82}, 038202 (2010).

\bibitem[Martinez~Torres et~al.(2008)]{MartinezTorres:2007sr}
A.~Martinez~Torres, K.~P. Khemchandani, and E.~Oset, \emph{Phys. Rev. C}
  \textbf{77}, 042203 (2008);
%
K.~P. Khemchandani, A.~Martinez~Torres, and E.~Oset, \emph{Eur. Phys. J.}
  \textbf{A37}, 233 (2008).

\bibitem[Xie et~al.(2010)]{Xie:2010ig}
J.-J. Xie, A.~Martinez~Torres, and E.~Oset  (2010), arXiv:1011.0852 [nucl-th].

\bibitem[Martinez~Torres et~al.(2009{\natexlab{b}})]{MartinezTorres:2009cw}
A.~Martinez~Torres, K.~P. Khemchandani, U.-G. Meissner, and E.~Oset, \emph{Eur.
  Phys. J.} \textbf{A41}, 361(2009{\natexlab{b}}). 

\bibitem[Albaladejo et~al.(2010)]{Albaladejo:2010tj}
M.~Albaladejo, J.~A. Oller, and L.~Roca, \emph{Phys. Rev.} \textbf{D82}, 094019
  (2010).

\bibitem[Martinez~Torres et~al.(2011)]{Torres:2011jt}
A.~Martinez~Torres, D.~Jido, and Y.~Kanada-En'yo, 
  arXiv:1102.1505 [nucl-th].

\bibitem[Kim et~al.(2011)]{Kim:2011ut}
K.I.~Kim, D.~Jido, and S.H. Lee, 
  arXiv:1103.0826 [nucl-th];
K.I.~Kim {\it et al.}, in this proceedings.

\bibitem[Nagahiro et~al.(2011)]{Nagahiro:2011jn}
H.~Nagahiro, K.~Nawa, S.~Ozaki, D.~Jido, and A.~Hosaka,
  arXiv:1101.3623 [hep-ph]; H.~Nagahiro, talk given in this conference. 

\bibitem[Cho et~al.(2010)]{Cho:2010db}
S.~Cho {\it et al.}  [ExHIC Collaboration], arXiv:1011.0852 [nucl-th];
A.~Ohnishi {\it et al.} in this proceedings.

\end{thebibliography}
%

\end{document}